\documentclass[aps,pra,twocolumn,showpacs,superscriptaddress,floatfix]{revtex4}
\usepackage{amsmath,amssymb,graphicx,graphics,color}
\usepackage{dcolumn}
\usepackage{bm}
\usepackage{psfrag}
\usepackage[T1]{fontenc}
\usepackage{calligra}

\begin{document}

\title{Topological Phases of Fermionic Ladders with Periodic Magnetic Fields}
\author{Gaoyong Sun}
\affiliation{Max Planck Institut f\"ur Physik komplexer Systeme, N\"othnitzer Stra\ss{}e 38, 01187 Dresden, Germany}

\begin{abstract}
In recent experiments bosonic [Atala et al., Nat. Phys. 10, 588 (2014),  B. K. Stuhl et al., Science 349, 1514 (2015)] 
as well as fermionic ladders [M. Mancini et al., Science 349, 1510 (2015)] with a uniform flux were studied 
and different interesting many-body states were observed. Motivated by these experiments, 
we extend the uniform synthetic magnetic field to a periodic case and show that a commensurate synthetic magnetic field 
offers an alternative scheme to realize topological phases in many-body systems of ultra-cold Fermi gases in ladder-like 
optical lattices. Using the exact diagonalization, we numerically determine the topological band structure, edge states, 
non-zero Chern numbers, Hofstadter-like-butterfly spectrum, and a complete phase diagram of non-interacting fermionic ladders.

\end{abstract}


\maketitle

\section{Introduction}
The effects of magnetic fields on quantum particles is of great interest, which already yielded the observation of many exotic quantum phases
in experiments, such as Hofstadter butterfly \cite{Hofstadter1976,Aidelsburger2013,Miyake2013} and
quantum Hall effects \cite{Klitzing1980,Tsui1982,Thouless1982} in two dimensional (2D) systems.
More exotic behavior, including spin liquid \cite{Sun2009,Chang2012}, bosonic integer quantum Hall effect \cite{He2015,Senthil2013,Furukawa2013,Liu2014},
density-dependent synthetic magnetism \cite{Greschner2015DM,Mishra1512} are proposed and engineered based on 
the uniform (or density-dependent) gauged fields \cite{Keilmann2011,Greschner2014,Greschner2015a}.
Ultra-cold atoms offer a clean and controllable platform to simulate these many-body systems \cite{Bloch2008,Lewenstein2007}.
The synthetic magnetic fields were recently generated to realize Hofstadter butterfly in ultra-cold gases
using laser assisted tunneling in 2D optical lattices \cite{Aidelsburger2013,Miyake2013}.
It is well known that 2D Hofstadter model can be mapped to one-dimensional (1D) Harper (or Aubry-Andr\'e) model \cite{Aubry1980,Lang2012,Kraus2012},
where periodic chemical potentials serve the same effect as magnetic fields.

Besides the realization of magnetic fields in 2D systems, chiral currents and vortex and Meissner phases were also observed
with an artificial magnetic field in bosonic ladders \cite{Atala2014}. 
Ladder systems with fluxes were also modelled experimentally with purely 1D but multi-component bosonic~\cite{Spielman}
as well as fermionic~\cite{Fallani} ultracold quantum gases, effectively realizing two-leg \cite{Piraud2015,Greschner2015,Hugel2014} 
and three-leg ladders \cite{Kolley2015}. 
In the latter two studies \cite{Spielman,Fallani}, the sites along the rung direction correspond to the different hyperfine states, 
forming a synthetic lattice dimension \cite{Celi2014,Zeng2015,Barbarino2015}. 
Interacting bosonic ladder systems with uniform flux besides the above mentioned Meissner 
and vortex phases that already exist for non-interacting particles, show various spontaneously symmetry broken phases \cite{Piraud2015,Greschner2015}.

There has been relatively little work on topological phases in ladder systems with synthetic magnetic field. Recently 
it was suggested that a ladder system in a uniform flux can mimic spin-orbit coupling
and reproduce edge modes of 2D Hofstadter model \cite{Hugel2014}. In another recent work shows that a 1/2 Laughlin-type 
fractional state can be realized in ladder systems with $\pi$ flux in every second plaquette \cite{Grusdt2014}. 
In addition, a triangular ladder-like geometry is exploited using trapped ions and atoms with next-nearest-neighbor interactions \cite{Grass2014},
showing that a uniform flux can split the dispersion and lead to topological phases, which are absent on a square ladder.


\begin{figure}
\includegraphics[width=8.6cm]{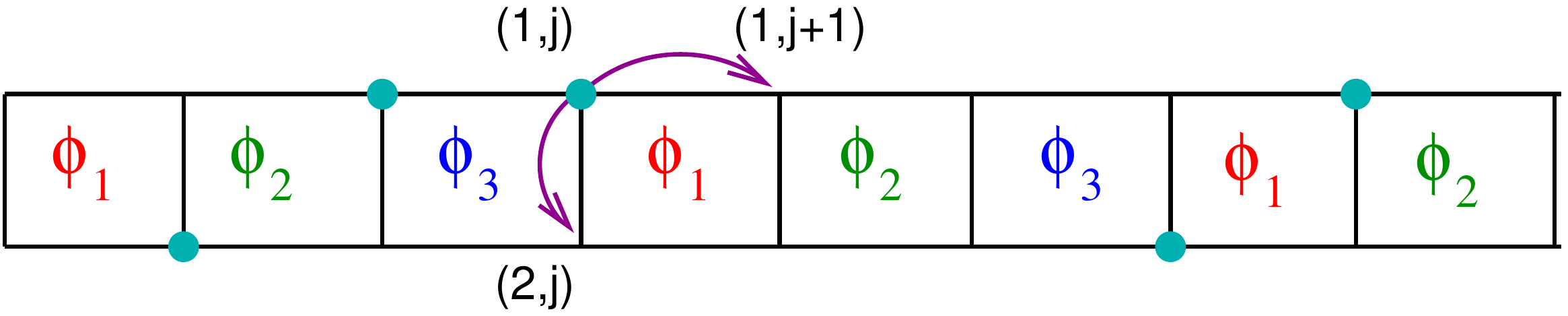}
\caption{(Color online) Two-leg ladder with periodic synthetic magnetic fields of periodicity three.
Particles can hop either along legs i.e. $(l,j)\rightarrow (l,j+1)$ or along rungs i.e. $(1,j)\rightarrow (2,j)$,
with the corresponding flux $\phi_{1}$, $\phi_{2}$, $\phi_{3}$ in neighbouring plaquette.
The total flux of a unit cell is $\Phi=\phi_{1}$ + $\phi_{2}$ + $\phi_{3}$.}
\label{lattice}
\end{figure}

In 1D systems, topological phases are usually introduced by a cosine-like modulation of either in chemical potentials \cite{Lang2012,Kraus2012,Deng2014}
or in hopping amplitudes \cite{Ganeshan2013}. One may also engineer topological phases by a simple extension of the above schemes
(or two-component systems with spin-orbit coupling \cite{Lin2011,Wang2012,Cheuk2012}) in ladders 
or in synthetic dimension lattices \cite{Spielman,Fallani,Zeng2015,Barbarino2015}.
In contrast to the above schemes, in this paper we focus on the effects of purely synthetic magnetic fields in ladder systems.
Using the exact diagonalization, we show that non-interacting fermionic ladders with periodic synthetic magnetic fields
can reproduce the whole Hofstadter-butterfly spectrum, non-trivial 2D topological band structures characterized by edge states and non-zero Chern numbers.
Hence, a ladder system with periodic synthetic magnetic fields can offer an alternative scheme for engineering topological states in many-body systems.

\begin{figure}[t]
\includegraphics[width=8.6cm]{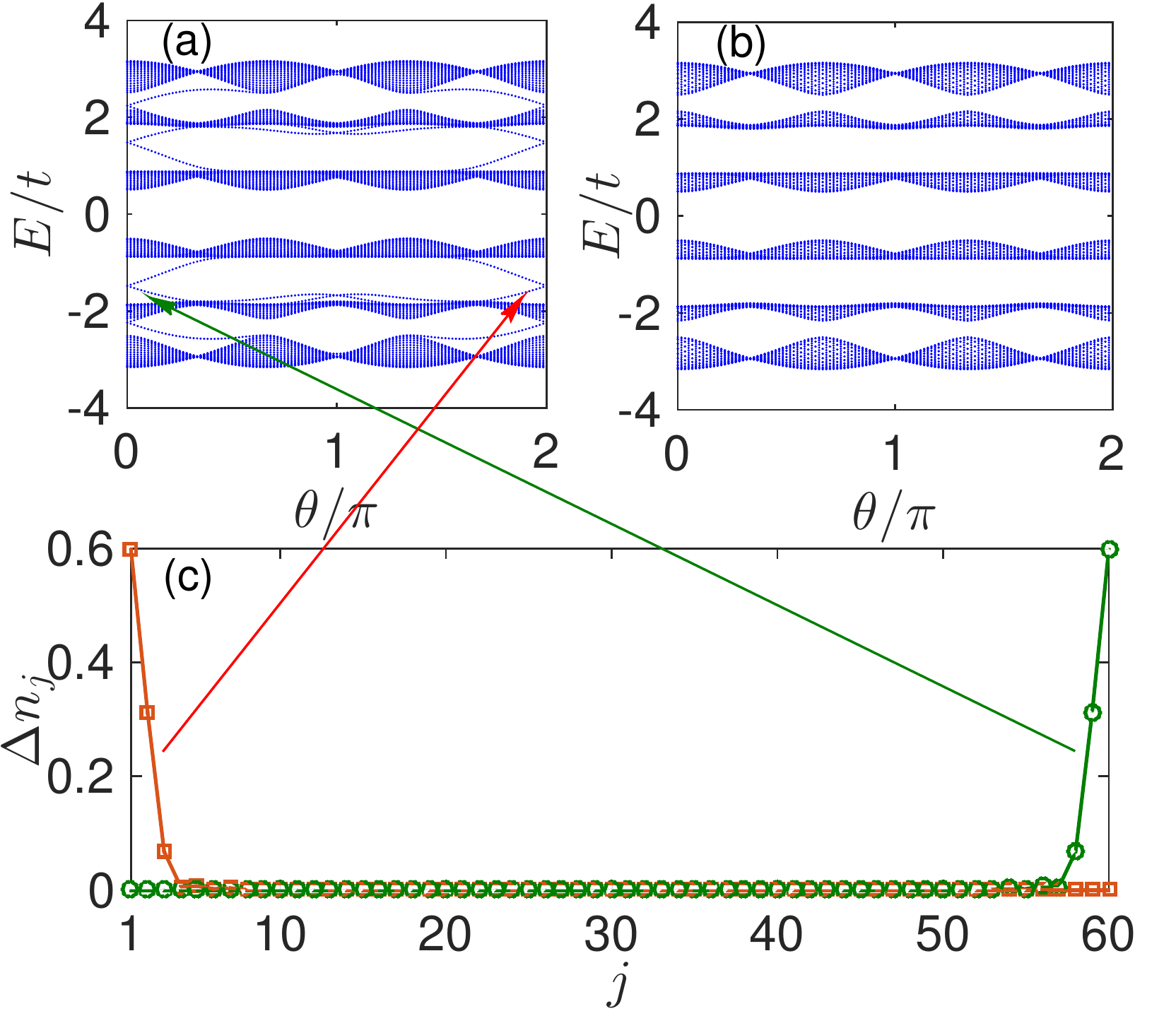}
\caption{(Color online) Energy spectra and edge states.
(a) Energy spectrum for $L=60$ rungs with $\alpha=1/3$ and $t_{\perp}=1.5$ in open boundary conditions.
(b) Energy spectrum with the same parameters of (a) in periodic boundary conditions.
(c) Edge states for $\theta=\pi/6$ and $\theta=11\pi/6$ respectively, denoted by arrows,
with $t_{\perp}=1.5$ around filling $\rho=1/3$ in open boundary conditions.}
\label{energy}
\end{figure}



\section{Model}
We consider a site-dependent hopping \cite{Atala2014,Spielman,Fallani,Zeng2015,Barbarino2015,Jaksch2003} system with $N$ fermionic particles 
in two-leg ladders shown in Fig.\ref{lattice},
\begin{align}
H =-{}& t\sum_{l=1,2; j=1}^{L} (e^{i(2 \pi \alpha j+\theta)f_{l}}c^{\dagger}_{l,j}c_{l,j+1}  + {\text H.c}) \nonumber \\
   -{}& t_{\perp}\sum_{j=1}^{L} (c^{\dagger}_{1,j}c_{2,j} +{\text H.c}) - \mu \sum_{l=1,2; j=1}^{L} n_{l,j},
\label{model}
\end{align}
where $(l,j)$ denotes a rung-chain index.
$c^{\dagger}_{l,j}$ ($c_{l,j}$) are creation (annihilation) operators at the $l$-th leg and the $j$-th site, $n_{l,j}=c^{\dagger}_{l,j}c_{l,j}$ 
is a density operator, and $\mu$ is the chemical potential.
$L$ denotes the total numbers of the rungs of two-leg ladder. The particle filling is given by $\rho=N/(2L)$.
Hoppings $t$ and $t_{\perp}$ are tunneling matrix elements along legs and rungs, respectively, 
and $\alpha=p/q$ is a commensurate number, with the $p$ and $q$ coprime integers, 
and $\theta$ an arbitrary phase.
We choose a gauge freedom and set $f_{1}=1$ for up-legs ($l=1$) and $f_{2}=0$ for down-legs ($l=2$) \cite{Note1}.
We use $t=1$ as an energy unit scale throughout the paper.

When $\alpha=1$, the flux ($\phi=\theta$) is uniform that has been realized and considered
in Refs. \cite{Atala2014,Fallani,Spielman,Piraud2015,Greschner2015,Hugel2014,Grass2014}.
When $\alpha=1/2$, it is a stagger flux ($\phi_{1}=\pi+\theta$, $\phi_{2}=2\pi+\theta$),
where the special case $\phi_{1}=0,\phi_{2}=\pi$ has been studied in Ref.\cite{Grusdt2014}.
When $\alpha=1/3$, the system has a periodic synthetic magnetic field of periodicity three
($\phi_{1}=2\pi/3+\theta$, $\phi_{2}=4\pi/3+\theta$, $\phi_{3}=2\pi+\theta$) shown in Fig.\ref{lattice}.
As a concrete example we will choose $\alpha=1/3$ (that is $p=1$, $q=3$) to show topological properties of a ladder throughout the paper \cite{Note2}.
It is important to note that the flux of a unit cell of the system
(with $\theta=0$) is $p \cdot (q+1) \cdot \pi$ that is an integer multiple of $\pi$.
Therefore the topological physics comes from the flux-introduced band spitting
similar to the 1D Aubry-Andr\'e model \cite{Lang2012,Kraus2012,Deng2014,Ganeshan2013}.

\begin{figure}[t]
\includegraphics[width=8.6cm]{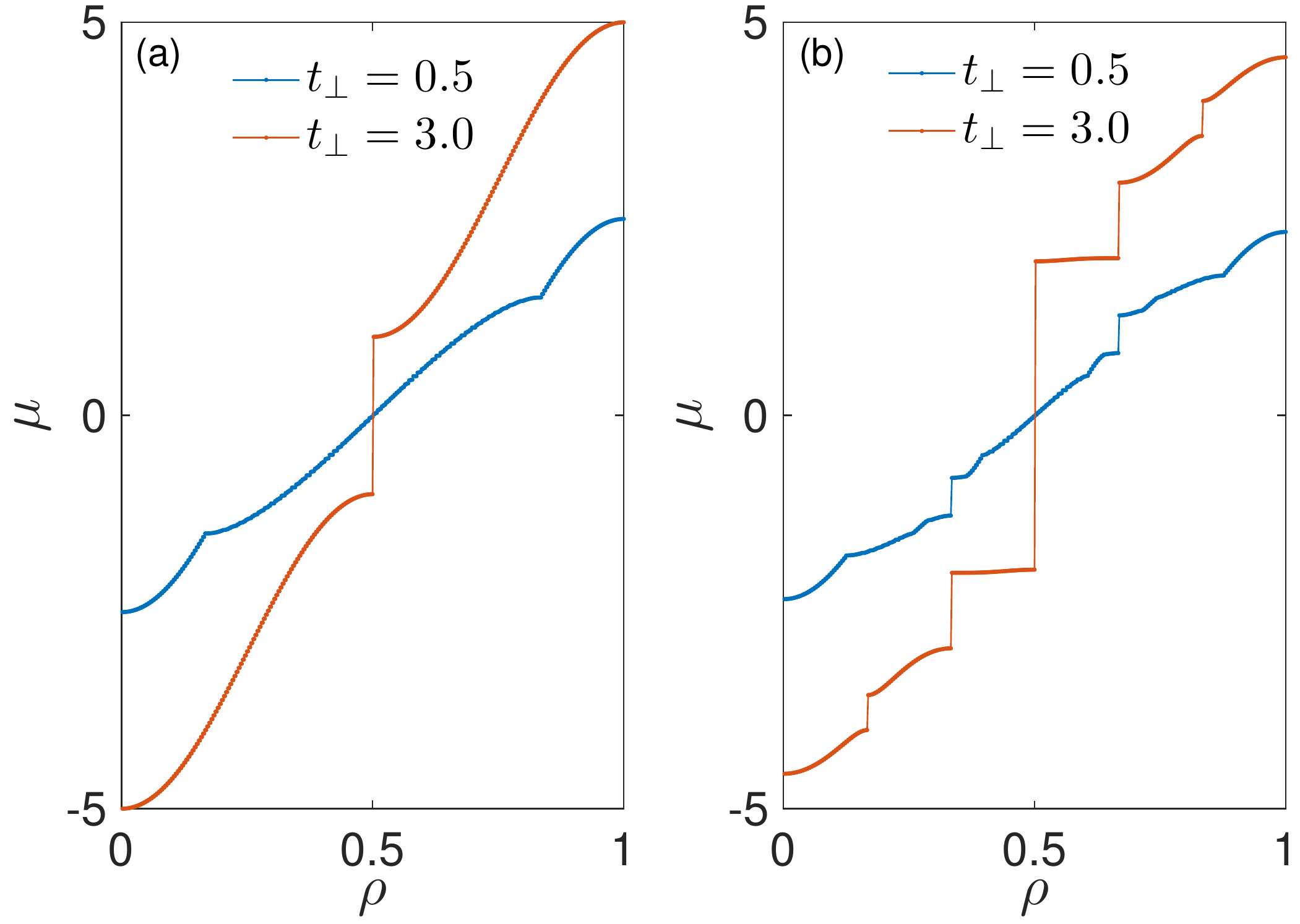}
\caption{(Color online) Chemical potential respect to density.
(a) Zero magnetic field $\alpha=0$, $\theta=0$ for $L=240$ rungs with $t_{\perp}=0.5$ (one band)
and $t_{\perp}=3$ (two bands) in periodic boundary conditions.
(b) Periodic magnetic field $\alpha=1/3$, $\theta=0$ for $L=240$ rungs with $t_{\perp}=0.5$ (three bands)
and $t_{\perp}=3$ (six bands) in periodic boundary conditions.}
\label{density}
\end{figure}

\section{Topological energy spectra}
\subsection{Energy spectra and the edge states}
The tight-binding model in Eq.(\ref{model}) can be exactly diagonalized in real space \cite{Lang2012,Kraus2012,Deng2014,Ganeshan2013,Leib1961}.
The system has usual two-band spectra in the absence of modulation, corresponding to the case $\alpha=0$. In presence of $\alpha=p/q$, 
each band is split into $q$ sub-bands, resulting in total of $2q$ bands. In order to see a 2D type band structure, 
we computed the band structure using $L=60$ rungs in the case of
$\alpha=1/3$ and $t_{\perp}=1.5$ by changing the phase $\theta$ from $0$ to $2\pi$
for both open boundary conditions (OBC) shown in Fig.\ref{energy}(a)
and periodic boundary conditions (PBC) shown in Fig.\ref{energy}(b).
It is clear that the band is split into 6 parts,
and all bands vary periodically with respect to $\theta$ with a periodicity $q$.

Note that the edge states appear only for OBC in contrast to PBC, confirming that we are dealing with a non-trivial topological phase.
The distribution of quasi-particles near filling $\rho=N/(2L)=1/3$ was computed in Ref. \cite{Zhu2013}
\begin{align}
 \Delta n_{j} = n_{j}(N+1)-n_{j}(N),
\end{align}
where $n_{j}(N)=\langle \Psi^{N}_{g} | \sum_{l}n_{l,j}| \Psi^{N}_{g}\rangle$ is the density profile of the ground state at filling $\rho=N/(2L)$.
As shown in Fig.\ref{energy}(c), edge states can appear either in a left side or in a right side depending on the phase $\theta$.
From band structure Fig.\ref{energy}(a), one can denote the following 4-type phases: (a) the system is a trivial
vacuum at filling $\rho=0$ and $\rho=1$; (b) the system is a band insulator at filling $\rho=1/2$;
(c) the system is in a topological phase when filling is $\rho=1/6,1/3,2/3,5/6$; (d) the system is a metallic state for all other fillings.
In contrast, for free fermions in the absence of flux, there are only 3 simple phases: vacuum, band insulator, and metallic state.

The corresponding $\rho-\mu$ relation is shown in Fig.\ref{density}, where one can find that:
in absence of magnetic field, it is a metallic phase with either four or two Fermi points \cite{Crepin2011} for $t_{\perp} < 2$.
Increasing $t_{\perp}$, the system starts to open a gap at $t_{\perp} = 2$, indicating that a band insulating phase is
starting to form at half filling $\rho=1/2$ for larger rungs hopping $t_{\perp} > 2$.
While in presence of periodic magnetic fields,
multi-bands appear separated by a gap, showing that the new topological phases emerge.

\subsection{Chern numbers}
To further confirm the topological properties of insulating phases, we calculate Chern numbers
defined in \cite{Lang2012,Deng2014,Zhu2013,Fukui2005,Varney2011} by introducing a twist angle $\varphi$ in hopping terms along legs
$t\rightarrow te^{i\varphi/L}$ to mimic 2D systems spanned by two momentum vectors $k_{x}$ and $k_{y}$.
\begin{align}
 C=\frac{i}{2\pi} \int_{0}^{2\pi} d\varphi \int_{0}^{2\pi} d\theta (\langle \partial_{\varphi} \Psi^{\ast}_{g} | \partial_{\theta} \Psi_{g} \rangle
 -\langle \partial_{\theta} \Psi^{\ast}_{g} | \partial_{\varphi} \Psi_{g} \rangle)
\end{align}

\begin{figure}[t]
\includegraphics[width=8.6cm]{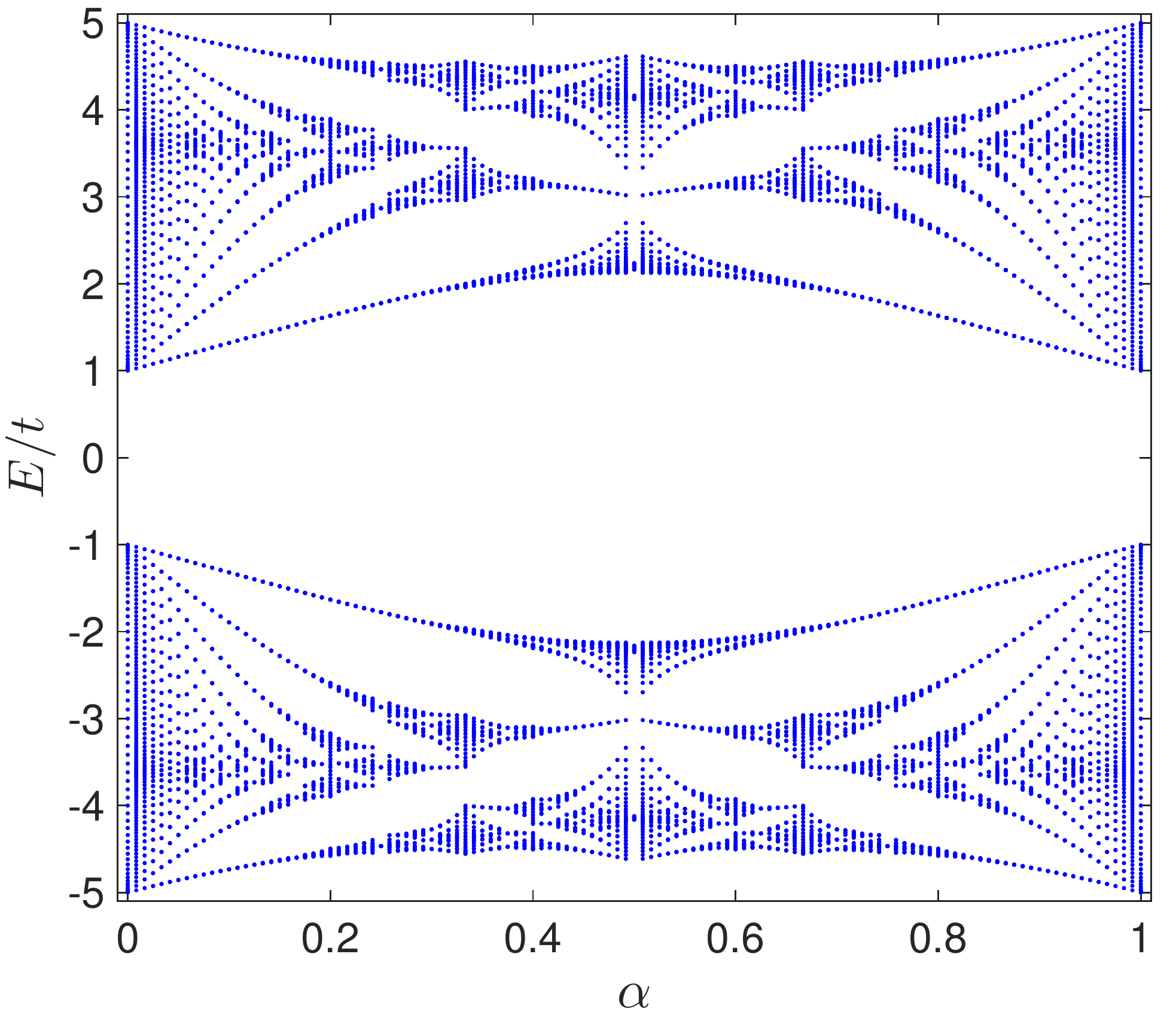}
\caption{(Color online) Fractal energy spectrum. The Hofstadter-butterfly-like energy spectrum as a function of $\alpha$
is obtained for $L=120$ rungs with $t_{\perp}=3$ and $\theta=0$ in periodic boundary conditions.
}
\label{Hofstadter}
\end{figure}

Numerical calculations from a discrete manifold \cite{Lang2012,Deng2014,Zhu2013,Fukui2005,Varney2011}
with PBC show that there are 3 different Chern numbers for different bands respectively:
(a) Chern number $C=0$ at filling $\rho=0$, $\rho=1/2$ and $\rho=1$;
(b) Chern number $C=1$ at filling $\rho=1/6$ and $\rho=5/6$;
(c) Chern number $C=-1$ at filling $\rho=1/3$ and $\rho=2/3$.
Non-zero Chern numbers indicate insulating phases at filling $\rho=1/6$, $\rho=1/3$, $\rho=2/3$ and $\rho=5/6$ are non-trivial topological phases.
The band structure and Chern numbers have the particle-hole symmetry (i.e. Chern number is $C=1$ for filling $\rho=1/6$ and $\rho=5/6$ )
with respect to half-filling $\rho=1/2$ because of the ladder geometry.

\begin{figure}[t]
\includegraphics[width=8.6cm]{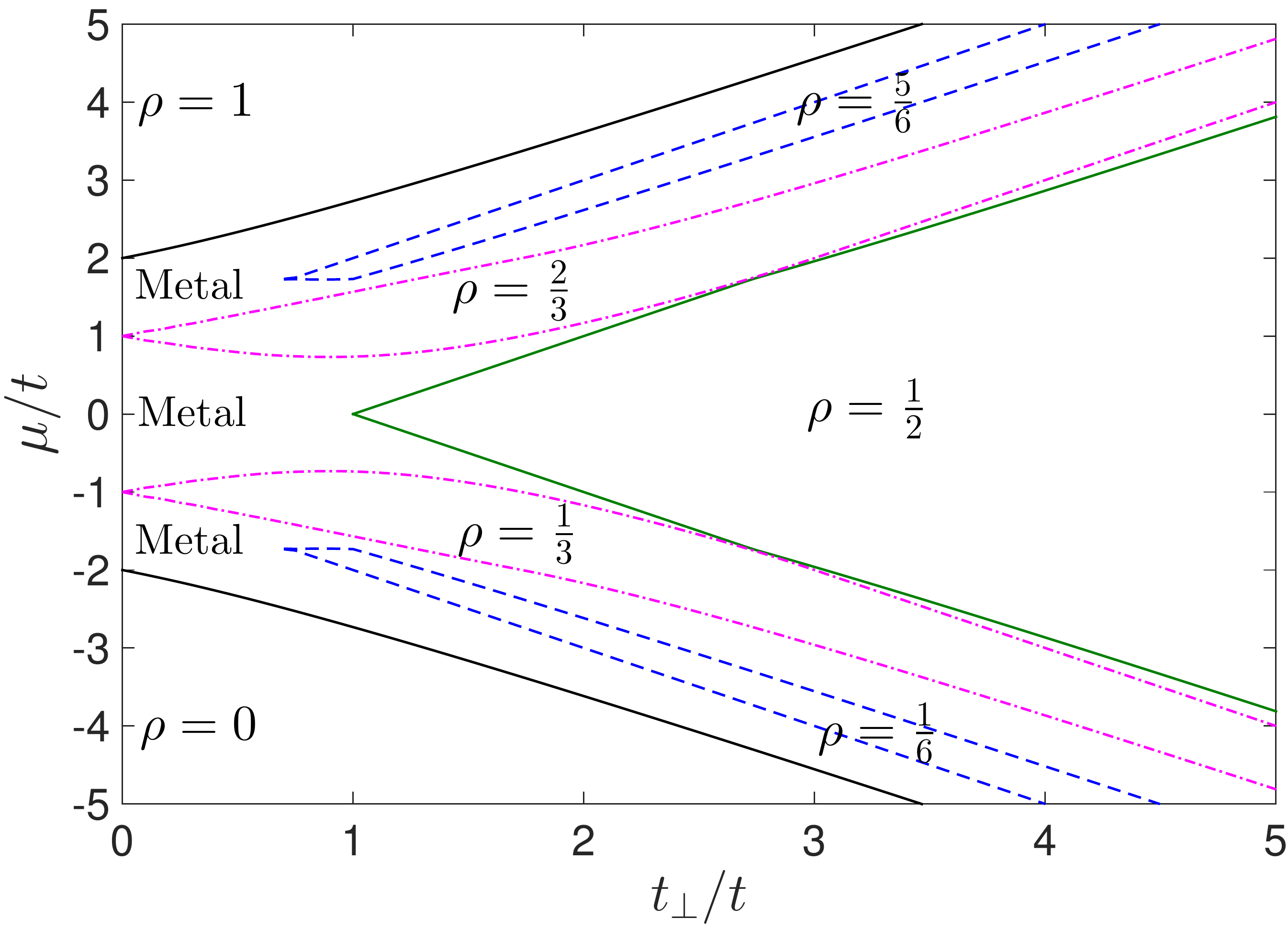}
\caption{(Color online) Phase diagram of non-interacting fermions for $L=120$ rungs with respect to $\mu$ and $t_{\perp}$
in the two-leg ladders with $\alpha=1/3$ and $\theta=0$ in periodic magnetic fields.
The density of metal phases are $0<\rho<\frac{1}{3}$, $\frac{1}{3}<\rho<\frac{2}{3}$ and $\frac{2}{3}<\rho<1$ respectively.
The insulating phases are denoted by the different densities. Where $\rho=0$ and $\rho=1$ are the vacuum states;
$\rho=\frac{1}{2}$ is the band insulator; $\rho=\frac{1}{6}$ and $\rho=\frac{5}{6}$ are the topological insulators with Chern number $C=1$;
and $\rho=\frac{1}{3}$ and $\rho=\frac{2}{3}$ are the topological insulators with Chern number $C=-1$.
}
\label{phasediagram}
\end{figure}

\subsection{Hofstadter-butterfly}
Another feature of such topological phases is fractional band structure. 
We diagonalize the system for $L=120$ rungs with $t_{\perp}=3$ and $\theta=0$
in PBC. As shown in Fig.\ref{Hofstadter}, there are two branches symmetric Hofstadter-like-butterflies that is
separated by a gap (minimal gap lying at $\alpha=0$ and $\alpha=1$) because of the effect of rungs hopping $t_{\perp}$.
We note that these two branching Hofstadter-butterflies are separated only when rung hoppings $t_{\perp} >2$
(see Fig.\ref{density}) since it is a metallic phase \cite{Crepin2011} for $t_{\perp} < 2$ at $\alpha=0$ and $\alpha=1$.

\section{Phase diagram}
To have a deeper understanding of topological phases, we present the whole phase diagram in Fig.\ref{phasediagram} in $\mu-t_{\perp}$ plane
using the exact diagonalization for $L=120$ rungs with $\alpha=1/3$ and auxiliary angle $\theta=0$.
From it, one can follow nicely how topological phases evolute.

For $J_{\perp}=0$, the ladder is decoupled into two 1D free fermionic chains,
for which magnetic fields can be gauged out. 
For $J_{\perp} \neq 0$ and zero flux $\alpha=0$, $\theta=0$, as mentioned before,
the system is a metallic state for $0<\rho<1/2$ and $1/2<\rho<1$ for any $t_{\perp}$. While at exactly half-filling $\rho=1/2$,
the system starts to develop a gap at $t_{\perp}/t = 2$ (at $\mu=0$) and become a band insulator.
In contrast to zero-flux ladder, a gap opens immediately for any infinitezimal 
$t_{\perp}$ with a periodic magnetic field $\alpha=1/3$ and $\theta=0$,
resulting in a topological phase with non-zero Chern number $C=-1$ at filling $\rho=1/3$ and $\rho=2/3$. 
The whole band is split into 3 subbands: band A ($\rho < 1/3$), band B ($1/3 < \rho < 2/3$) and band C ($\rho > 2/3$), 
separated by the filling $\rho=1/3$ and $\rho=2/3$. 
Further increasing the tunneling $J_{\perp}$,
band A and band C each start to split into two subbands resulting in a topological phase 
with Chern number $C=1$ at filling $\rho=1/6$ and $\rho=5/6$.
When $t_{\perp}/t = 1$ (at $\mu=0$ at $\rho=1/2$ ), the last band B gets starting to split,
which happens earlier compared to the zero-flux free fermionic case ($t_{\perp}/t=2$).

In summary, a ladder in a periodic magnetic field has more chance to form insulating phase 
because of the special band structure, preventing particles hopping from one subband to another.

\section{Conclusion and outlook}

We have shown that a ladder system can present topological phases and Hofstadter-butterfly-like energy spectrum
with a periodic synthetic flux.
The periodic synthetic magnetic field plays a similar role as superlattices
in contrast to the uniform flux that induces a Meissner phase and a vortex phase.
Edge states and non-zero Chern numbers verify that insulating phases are non-trial topological phases.

Our study opens new questions: whether a ladder system with periodic flux can exhibit all similar physics
as 1D chain model with superlattices. It would be very interesting to extend the system
to interacting systems to check whether ladder systems with the periodic flux and long-range interactions
can also give rise to a fractional topological phase \cite{Xu2013}, to check whether a many body localization phase
can exist in interacting ladders when $\alpha$ is an irrational number
(or when systems are placed in disordered magnetic fields) \cite{Li2015}.

It would be also very interesting to understand the physics of 
interacting hard-core bosonic and soft-core bosonic systems,
where new phases are expected to appear as is the case for uniform flux \cite{Crepin2011,Piraud2015,Greschner2015}.
Such issues are beyond the scope of this paper, which is left for future investigation.

\section{Acknowledgments}
G. S would like to thank L. Santos, X. Deng, S. Greschner, T. Pohl, A. Eckardt
for useful discussions, and especially thank T. Vekua for a careful reading of manuscript and useful suggestions for the paper.
G. S would like to thank the support of QUEST (Center for Quantum Engineering and Space-Time Research)
and Institute for Theoretical Physics of Leibniz Universit\"at Hannover, where part of the work was done
and Visitors Program of Max Planck Institute for the Physics of Complex Systems.

\end{document}